\documentclass[conference]{IEEEtran}
\usepackage[pdftex]{graphicx}
\usepackage{dsfont}
\usepackage{braket}

\begin{document}
\title{The Effective Solving of the Tasks from NP by a Quantum Computer}

\author{\IEEEauthorblockN{Sergey Sysoev}
\IEEEauthorblockA{Dep. of Mathematics anf Mechanics\\
St.Petersburg State University\\
St.Petersburg, Russia \\
Email: sysoev@petroms.ru}}

\maketitle

\begin{abstract}
The new model of quantum computation is proposed, for which an effective algorithm for solving any task in NP is described. The work is based on and inspired by the Grover's 
algorithm for solving NP-tasks with quadratic speedup compared to the classical computation model. The provided model and algorithm exhibit the exponential speedup over that described by Grover. The main idea of the model improvement is employing the von Neumann architecture for a quantum computer, allowing program parts to be stored as quantum states along with the processed data.   
\end{abstract}


\IEEEpeerreviewmaketitle

\section{Introduction}
The mathematical model of quantum computation has appeared in the 1980-th with the works of David Deutsch [1] and others [2-4]. 
In 1994 Peter Shor [5] proposed an effective quantum algorithm for solving the NPI-candidate task -- factorisation of big composite numbers. This breakthrough allowed to expect, 
that any task from NP could have an effective solution on a quantum computer.
This hope was weakened after the Lov Grover's work [6] in 1996, which proposed the general quantum algorithm for solving any task from NP. 
The algorithm was designed to search the particular point in the unsorted database of size N with only $\sqrt{N}$ oracle queries. 
Grover has shown that under the considered assumptions the algorithm is optimal, so only quadratic speedup can be achived by the quantum computer compared to the classical one, 
if we don’t have additional information about the oracle functioning.

This paper describes the improvement of the quantum database search algorithm, exhibiting the exponential speedup over the classical search, by changing some preliminary assumptions about the model of computation.

The reasoning goes as follows. In Section 2 we recall the Grover's database search algorithm. Section 3 describes the new quantum computation model with von Neumann Architecture applied, 
and the new algorithm for the unsorted database search is proposed along with its complexity estimation. Section 4 glues everything together thus serving as a conclusion.

\section{The Grover's Algorithm}

First, let us recall the algorithm proposed by Lov Grover. 
Let's consider the function f:

$$ f : {\{0,1\}}^n \rightarrow {0,1} $$
$$ \exists ! \omega : f(x) = 1 \iff x = \omega, $$

which is some decision function over the set of n-bit strings. 
$N = 2^n$ --  is the number of all possible n-bit strings, thus the number of all possible inputs to f.
Function f is implemented as a black box. Our purpose is to find the n-bit string $\omega$ on which f returns 1.
Grover defines the quantum oracle $U_{\omega}$:

$$U_{\omega} : \ket{x}\ket{y} \rightarrow \ket{x}\ket{y \oplus f(x)}, $$

where x is n-bit string, and y is one bit. 
On the first n qubits $U_{\omega}$ acts as follows:

$$ U_{\omega} = \mathds{1} - 2 \ket{\omega}\bra{\omega}. $$

$U_{\omega}$ acts identically on any vector orthogonal to $\ket{\omega}$, while changes the sign of $\ket{\omega}$ itself, 
thus it can be considered as the reflection of any vector over the hyperplane orthogonal to $\ket{\omega}$.

Then the vector $\ket{s}$ and operator $U_s$ are introduced as follows:

$$ s = H{\ket{0}}^n, $$
$$ U_s = 2\ket{s}\bra{s} - \mathds{1}, $$

where H is n-qubit Hadamard transform. Operator $U_s$ reflects any vector over the vector $\ket{s}$. 
The Grover Iteration is

$$ U_{grov} = U_s U_{\omega}. $$

\begin{figure}[!t]
\label{grov_1}
\centering
\includegraphics[width=2.5in]{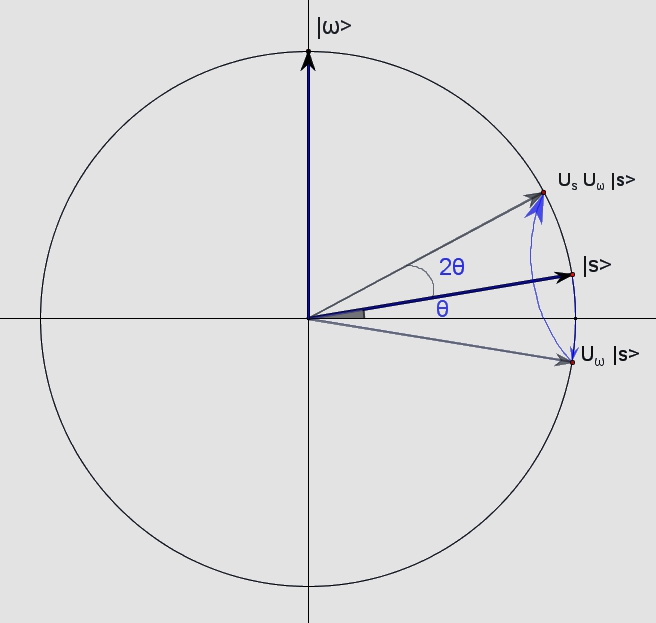}
\caption{The First Grover Iteration.}
\end{figure}

\begin{figure}[!t]
\label{grov_2}
\centering
\includegraphics[width=2.5in]{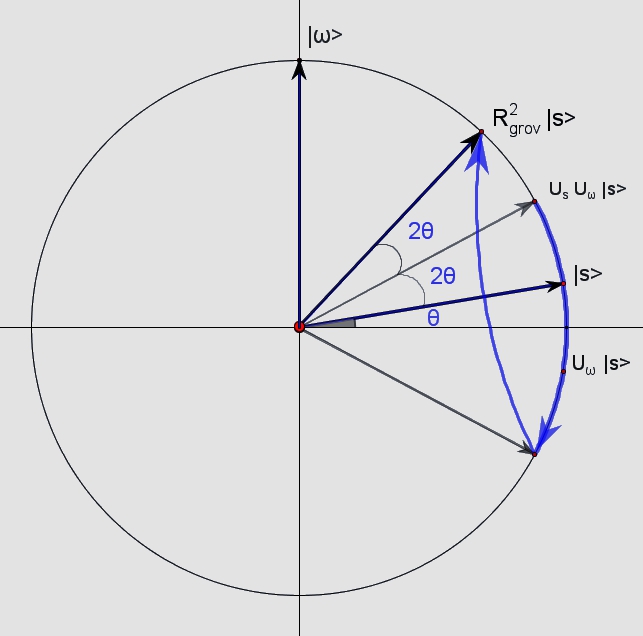}
\caption{The Second Grover Iteration.}
\end{figure}

The $U_{grov}$ operator rotates the initial vector $\ket{s}$ towards the desired vector $\ket{\omega}$ by the angle $2 \theta$, where $\sin{\theta} = \frac{1}{\sqrt{N}}$. 
The action of the first two Grover iterations is shown on the figures \ref{grov_1} and \ref{grov_2}. 

\section{The Algorithm}

The von Neumann Architecture was introduced by John von Neumann [7] in 1945. One of its most famous concepts is the idea of representing algorithms as a special type of data, which can be stored in computer memory along 
with the information they process. The computer hardware may then become universal and not depending on the solved tasks. 

This idea is considered to be very common and natural for the classical computing model. However, the quantum computing model still distinguish data (quantum system state) from a program (sequence of unitary transforms in the 
system state space). Some of these unitary tranforms have the natural mapping on the system states. For example, the state $\ket{s}$ defines the transform $U_s = 2\ket{s}\bra{s} - \mathds{1}$ in the previous section. 

If we assume, that the state of one quantum system can be used as a reflection tranform in another quantum system, then we can introduce a much simplier and effective algorithm for the unsorted database search.

Following Grover, we will denote:

$$ f : {\{0,1\}}^n \rightarrow {0,1}, $$ 
$$U_{\omega} : \ket{x}\ket{y} \rightarrow \ket{x}\ket{y \oplus f(x)}, $$
$$ \ket{s} = H \ket{0}^n. $$

\begin{figure}[!t]
\label{new_alg_1}
\centering
\includegraphics[width=3.5in]{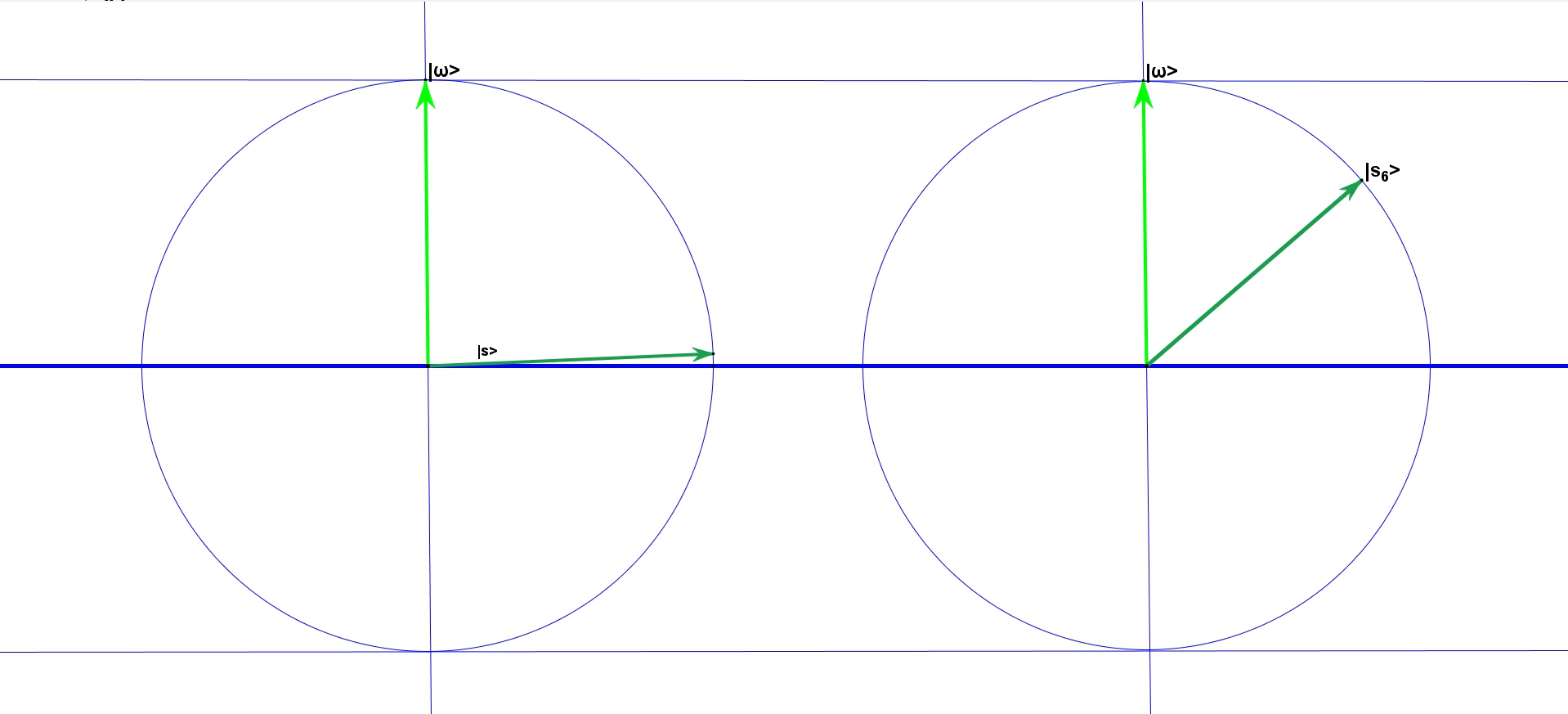}
\caption{The left system is in the initial state, while the right system is on the half of its way to $\omega$ with 6 Grover's iterations.}
\end{figure}

We still consider the case of searching of the special input $\omega$ among all n-bit strings. The number of Grover's iterations needed to get the result with high probability after measurement is estimated by 
$T_{grov} = \frac{\pi\sqrt{2^n}}{4}$. For example, if n equals to 8 (each input is one byte), then the number of Grover's iterations must be close to 12. 
Let's suppose that we already have performed a half of these 12 iterations in one system, while in another system we've prepared the initial state $\ket{s}$ (figure \ref{new_alg_1}).

\begin{figure}[!t]
\label{new_alg_2}
\centering
\includegraphics[width=3.5in]{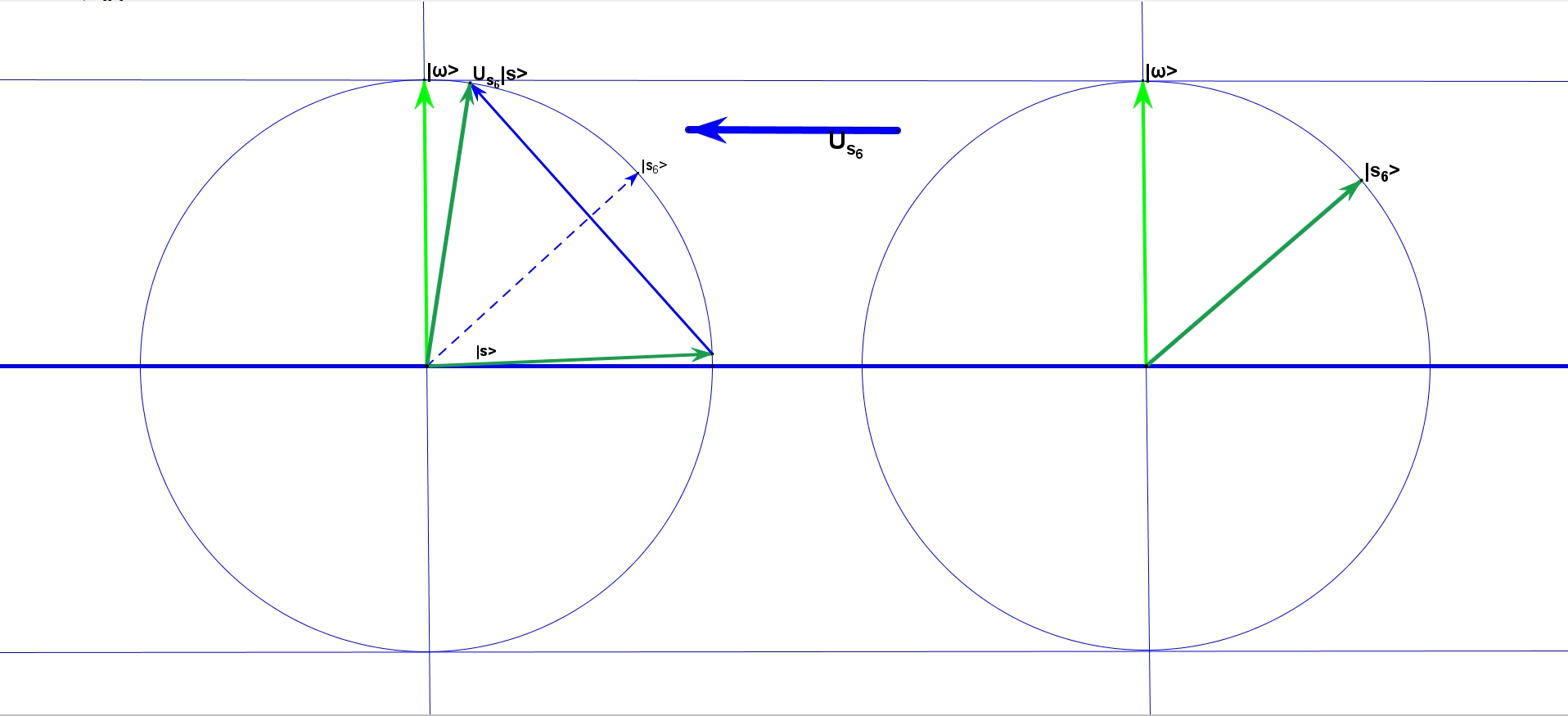}
\caption{$U_{s_6}$ applied.}
\end{figure}

If the vector $\ket{s_6}$ can be used as the operator $U_{s_6} = 2\ket{s_6}\bra{s_6} - \mathds{1}$ in the left system with the initial state $\ket{s}$, then we need only one iteration to get close to $\omega$ instead of 
the remained 6 Grover's iterations (figure \ref{new_alg_2}).

\begin{figure}[!t]
\label{new_alg_3}
\centering
\includegraphics[width=3.5in]{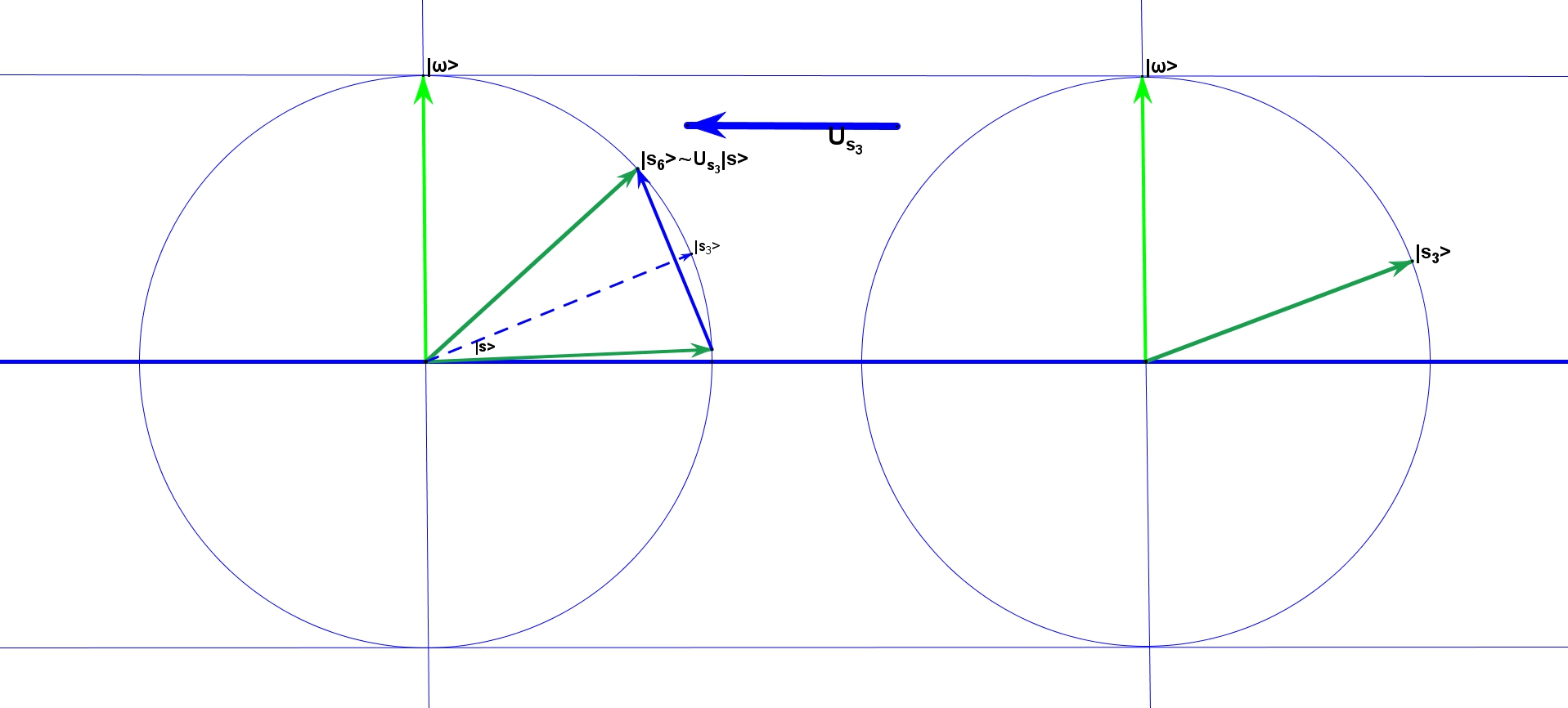}
\caption{$U_{s_3}$ applied.}
\end{figure}

Again, the vector $\ket{s_6}$ itself can be constructed from the vector $\ket{s_3}$ with the same process, thus omitting 3 Grover's iterations (figure \ref{new_alg_3}). 
Continuously applying this trick we can reduce the number of all Grover's iterations to $T_{new} = \lg{T_{grov}}$. Each such reflection divides the number of nesessary Grover's iterations by 2.

This means that for solving the unsorted database search problem we need to:
\begin{enumerate}
	\item{perform one Grover's iteration to get vector $\ket{s_1}$.}
	\item{prepare the state $\ket{s}$ on another system.}
	\item{apply the vector from the first system as a reflection to the second system.}
	\item{repeat $\lg{T_{grov}}$ times the steps 2 - 3 .}
\end{enumerate}
The implementation of $U_{s_i}$ operator, which depends on some quantum state $\ket{s_i}$ and acts on another quantum system is an open question in the scope of this work.

\section{Conclusion}

The proposed quantum algorithm allows us to solve the unsorted database search problem exponentialy faster than that described by Grover. However, the assumption of possibility of applying states as operators to other
systems doesn't correspond to the classical quantum computation model, developed to these days.  

Note that with the new algorithm we need only one call to the $U_{\omega}$ oracle to get the direction. After that all the remained iterations do not call the oracle directly.
To implement the new algorithm we need only two quantum systems, which will change their roles on each iteration. 
The "algorithm" system (which is used as an operator) on the next iteration must prepare the initial state $\ket{s}$ and become a "data" system and vice versa.

\end{document}